# Airborne dispersion of droplets during coughing: a physical model of viral transmission

Hongying Li, PhD[1], Fong Yew Leong, PhD[1†], George Xu, PhD[1], Chang Wei Kang, PhD[1], Keng Hui Lim, PhD[1], Ban Hock Tan, FRCP(UK)[2], Chian Min Loo, MBBS[3]

[1] A*STAR Institute of High Performance Computing, 1 Fusionopolis Way, Connexis, Singapore 138632

[2] Department of Infectious Diseases, Singapore General Hospital, Outram Road, Singapore 169608

[3] Department of Respiratory and Critical Care Medicine, Singapore General Hospital, Outram Road, Singapore 169608

[†] Correspondence: F.Y. Leong, Ph.D., A*STAR Institute of High Performance Computing, 1 Fusionopolis Way, Connexis, Singapore 138632. Phone: (65) 6419 1531. Email: leongfy@ihpc.a-star.edu.sg

**Abstract**

The Covid-19 pandemic has focused attention on airborne transmission of viruses. Using realistic air flow simulation, we model droplet dispersion from coughing and study the transmission risk related to SARS-CoV-2. Although most airborne droplets are 8–16 μm in diameter, the droplets with the highest transmission potential are, in fact, 32–40 μm. Use of face masks is therefore recommended for both personal and social protection. We found social distancing effective at reducing transmission potential across all droplet sizes. However, the presence of a human body 1 m away modifies the aerodynamics so that downstream droplet dispersion is enhanced, which has implications on safe distancing in queues. Based on median viral load, we found that an average of 0·55 viral copies is inhaled at 1 m distance per cough. Droplet evaporation results in significant reduction in droplet counts, but airborne transmission remains possible even under low humidity conditions.

**INTRODUCTION**

The current coronavirus disease outbreak[1] is an unprecedented global crisis with confirmed cases in the millions. The exceptional infectiousness of the severe acute respiratory syndrome coronavirus 2 (SARS-CoV-2) has focused attention[2] on the nature of its transmission pathways, with suspicion for an airborne route.[3–5] Airborne transmission depends on three main factors, namely, stability of virus, air circulation and droplet deposition.[6] SARS-CoV-2 is found to be stable in aerosol for up to the three hours of an experiment[7] similar to its predecessor SARS-CoV-1 which in certain circumstances, achieved airborne transmission.[8] Airborne transmission of SARS-CoV-2 cannot be dismissed even though epidemiological studies giving an R0 between 1 and 2 do not support airborne transmission as the major route of transmission.[9] Recent air sampling conducted in airborne infection isolation rooms has found PCR-positive particles of sizes in excess of 1 μm despite extensive air changes.[5]

During a cough or sneeze, mucosalivary fluid is expelled into the surrounding air in the form of droplets.[10,11] Droplets greater than 5 μm in diameter are termed as respiratory droplets whereas those less than 5 μm in diameter are droplet nuclei.[12] The size of the droplets affects the range of dispersal significantly.[13] Respiratory droplets tend to settle quickly and contaminate surrounding surfaces within a short distance,[6,14] whereas droplet nuclei can remain airborne for hours and present a long-range transmission risk.[15]

As a first line of defence against the pandemic, many countries have adopted what is commonly known as 'social distancing' where individuals are advised, sometimes legally mandated to maintain a certain distance from other individuals in public. Mathematical models[16] suggest that enforced physical separation could be an effective measure when deployed swiftly during a viral outbreak,[17] based also on simulations of past viral outbreaks.[18] The actual recommended distance varies from 1 m (Singapore), 1.5 m (Australia), 6 feet (CDC, USA) to 2 m (UK).

Generally public guidelines range from 1 to 3 m, and the 'science behind these numbers', as policy makers put it, is loosely based on an early seminal work.[19] In a recent study based on turbulent cloud physics, cough droplets are reported to spread up to 7–8 m.[20] Further, researchers at Wuhan hospitals found corona-virus residues in floor samples up to 4 m from identified sources.[21] Hence, even though the rationale behind social distancing is robust,[22] there is clearly no consensus as to what constitutes a safe separation distance, even for health workers treating infected patients.[23]

Fluid dynamics plays an important role in almost every aspect of this pandemic.[24] A brief survey of cough dispersion studies[25] yields theoretical puff model[11] and plume model[26], supported by visualization techniques such as schlieren[27], shadowgraph[28], and particle image velocimetry[29,30]. Notably, numerical methods, such as Computational Fluid Dynamics (CFD) based on Reynolds Averaged Navier-Stokes (RANS) turbulence models[31] produce high resolution flow fields and concentration data,[32] which not only compensate for slow instrumental speeds of analytical techniques,[25] but are also adaptable to different environments and scenarios, such as passengers in an aircraft cabin,[33] and more recently, a cough dispersion study in an outdoor environment under significant wind speeds,[34] whose results are useful in integrated transmission modeling.[35]

It is important to note that droplet dispersion model may infer transmissibility but not the actual infection risk. That will take an infectious dose, which refers to the number of viral particles to establish an infection in half of individuals, which depending on the type of virus.[36] Risk assessment of airborne infection should also account for actual viral transport.

In this study, we performed numerical simulations of droplet dispersion from a single cough based on single person and two person settings under realistic indoor conditions and assess viral transmission through airborne droplets.

**METHODS**

**Droplet dispersion model**

We consider a standing person who initiates a sudden, involuntary cough in an indoor environment. For reference we shall label the person the 'Cougher'. Our objective here is to assess the droplet dispersion potential under representative conditions: slight breeze from behind towards the Cougher at a speed 0·3 m/s, an ambient air temperature of 25 °C and a relative humidity of 60% (typical humidity in an air-conditioned environment). As detailed in the Supplementary Information, the model cough is inclined downwards at an average of 27·5°,[37] follows a characteristic air flow pattern[33,37] at breath temperature of 36°C, and emits a cluster of droplets with a standard size distribution[11,38] and viral loading[39]. The Cougher model is based on a standard human 1·7 m tall (average height for a Singaporean male) with open arms who begins normal breathing cycles immediately after the cough.

Separately, we consider the case of two persons, with the other labelled the 'Listener'. The Listener model, positioned either 1 or 2 meters away from the Cougher, is based on a standard human 1·59 m tall (average height for a Singaporean female) with arms closed and she begins normal breathing cycles immediately at the start of the cough.

Using Eulerian-Lagrangian model, we solve numerically for Newtonian air flow, temperature, species and droplet trajectories in space and time, using finite volume method implemented on ANSYS FLUENT (2019R3). First, steady state Eulerian simulations with species transport are performed for humid air flow and heat transport within the domain. Then using steady state solutions as input, simulations are repeated using discrete phase model, where transient momentum and heat transfer are included in Lagrangian tracking of droplet trajectories.

Droplets are assumed stable and do not experience breakup. The governing equations and computational details are available in the Supplementary Information.

**Viral transmission analysis**

We collect all droplets past designated distances, specifically 0·5, 1·0, 1·5 and 2·0 m from the source and characterize their respective size distributions time-averaged up to 10 s since the onset of the cough, and adjusted for wind speed (0·3 m/s). The droplet count probability is a probability function obtained by taking the ratio of the number of droplets of a certain size and the width of the corresponding size interval based on the original size histogram.[38] To assess viral transmission potential, we calculate the number of viral copies found in each droplet in terms of the viral load (copies per volume). The median viral count probability is a probability distribution obtained by taking the product of median SARS-CoV-2 viral load ($3·3 \times 10^6$ copies/mL)[39] and the droplet count probability distribution. The viral loading found in saliva is assumed to be homogeneously distributed among droplets emitted during a cough.

**Risk assessment**

To assess the extent of exposure, we collect droplets deposited on the Listener model surfaces, over the entire body (surface area 1·43 $m^2$) and the mouth region (3·5 $cm^2$), assuming mouth breathing, and including normal breathing cycles. Separately, we include a scenario where the Listener is wearing a mask (aspect ratio 1·43) and account for droplets landing on the mask (288 $cm^2$). The droplet deposition count probability is a probability function obtained by taking the ratio of the number of deposited droplets of a certain size and the interval of that size based on the original size histogram.[38] The median viral deposition count probability is a probability distribution obtained by taking the product of median SARS-CoV-2 viral load ($3\cdot3\times10^6$ copies/mL)[39] and the droplet deposition count probability distribution. Total deposited viral counts are obtained by integrating the respective median viral count probability distributions within size distribution limits (2–100 μm). Pulmonary deposition efficiency is not considered, since all droplets are in micron size range and do not escape readily via exhalation.[40] In addition, we do not distinguish between exposure to ingested or inhaled droplets, and trapped droplets do not re-entrain.

**RESULTS**

**Droplet dispersion**

For the base case with a single Cougher, Figure 1 shows snapshots of 2–100 μm droplet dispersion up to 10 s following a cough, side and top-down view, without (top) and with evaporation physics (bottom). Generally, larger droplets (in red) separate from the cloud and settle in seconds with a dispersion range of barely 1 m; smaller droplets (in blue) are buoyant and spread over large distances but may also evaporate rapidly into residues known as droplet nuclei.[19] For non-evaporative case (Figure 1; top), the droplet cloud starts off as a fast moving puff[11] from the oral region but quickly disperses into a plume[26] inclined at an angle from 14° to 10° (10 s) from the chest. Top-down view shows that the lateral droplet dispersion in depth is relatively constant, and by inspection, the plume can be effectively confined within a 20° forward wedge. Droplets near the sides of the virtual wedge travel faster than those near the center due to entrainment of air flow into the wake.

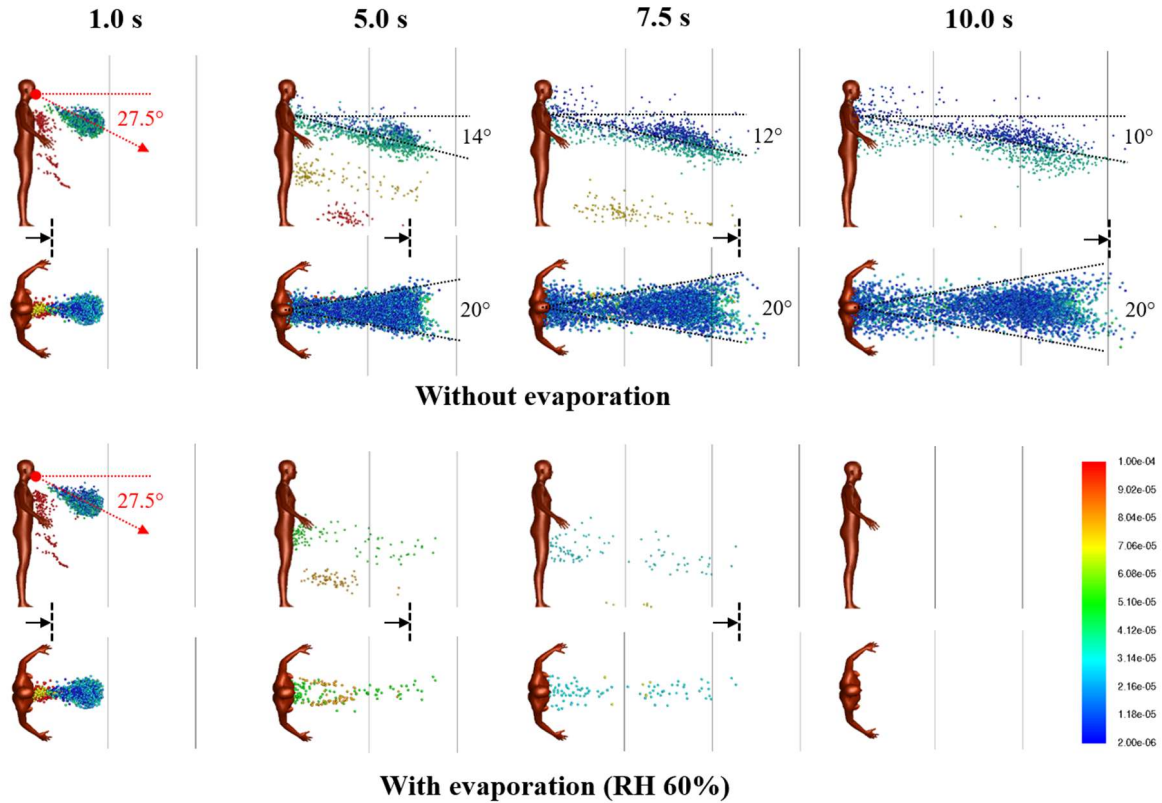

**Figure 1.** Droplet dispersion (side and top-down views) from a single cough inclined downwards at 27·5° for non-evaporative (top) and evaporative (bottom) cases at relative humidity of 60%. Color bar indicates droplet sizes (2–100 μm). Vertical lines are spaced 1 m apart and arrows are drift markers based on a background wind speed of 0·3 m/s. Ambient air temperature is 25°C and breath temperature is 36°C. Plume angles 14–10° from the chest 10 s after the cough and lateral dispersion fits a 20° forward wedge.

For the evaporative case, significant reduction in droplet counts is observed seconds from the cough (Figure 1; bottom). At relative humidity of 60%, the lifetime of a droplet 10–100 μm is in the order of seconds (Supplementary Figure S2). Small droplets rapidly evaporate into droplet nuclei, which continue to remain airborne for a long time due to their small size. As evaporating droplets decrease in size, the settling times are increased, resulting in a horizontal plume at hip level. Droplets are also observed to become more mono-disperse (a decrease in diversity of droplet sizes) compared to non-evaporative case. Top-down view shows minimal lateral dispersion for the remaining droplets. Taking the 20° forward wedge plume from the non-evaporative case, it is likely that evaporated droplets, or droplet nuclei, would follow similar dispersion trends.

The droplet count probability depends on the droplet size distribution and distance from the Cougher (Figure 2). The mode of the distribution lies between 8 and 16 µm; closer to 8 µm at short distances (<1 m) but closer to 16 µm at longer distances. Increase in distance results in reduction of droplet counts across all droplet sizes. Interestingly, we find that droplets between 32 and 40 µm represent the highest transmission potential in terms of droplet sizes. Fewer in numbers compared to 8–16 µm droplets, these 32–40 µm droplets contain much higher viral counts due to their larger volumes. Larger droplets greater than 75 µm contain even higher viral count per droplet, but they tend to settle rapidly and therefore present little airborne transmission potential except under strong wind conditions.[34]

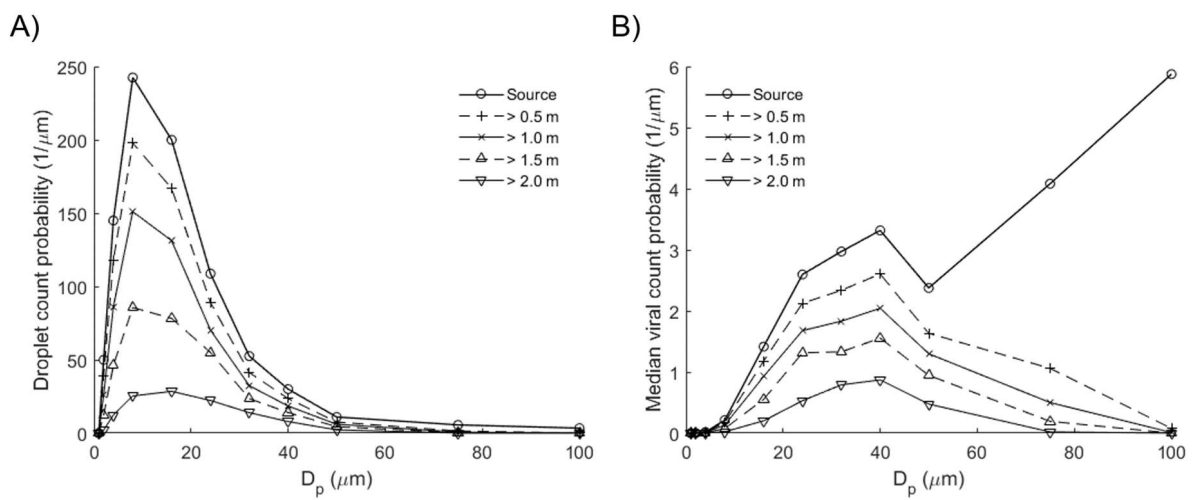

**Figure 2**. Airborne droplets with horizontal distances exceeding indicated distances from source, time-averaged up to 10 s since onset of cough, adjusted for wind drift (0·3 m/s). A) Droplet count probability. Mode: 8–16 μm. B) Median viral count probability based on median viral load (3·3×10$^6$ copies/mL)[39]. Mode: 32–40 μm. Droplets larger than 75 μm have low airborne transmission potential despite high viral counts.

Exposure to virus via droplets depends on distancing, as well as wind speed and direction (Figure 3). For 0·5 m distancing without evaporation, the median viral count accumulates rapidly to 80 in less than 1 s due to high velocity droplet transport from the cough jet, which is substantially faster than the background wind drift. Between 5 to 9 s, viral counts are elevated from 100 copies up to 147 due to the passage of a cloud of large droplets drifting close to the ground before settling (Figure 1; 5–7.5 s); these droplets contain high viral counts. As distancing increases, the droplets become more wind dispersed, in this case, at a wind speed of 0·3 m/s. In particular, it takes approximately 7 s for viruses to be projected over a distance of 2 m. This delay may provide sufficient time for simple reactive measures, such as stepping away or wearing a surgical mask.

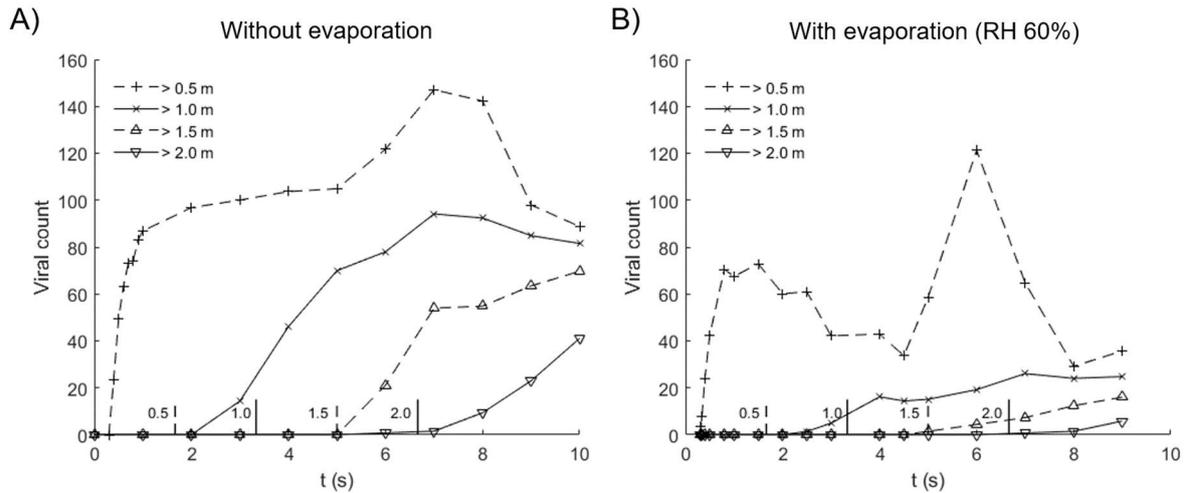

**Figure 3.** Median viral count collected at indicated distances from source in time, for cases A) without evaporation and B) with evaporation at relative humidity of 60%. Evaporated droplets persist as droplet nuclei which may pose infection risk over extended distance and time scales. Vertical lines along the abscissa denote the time required for wind drift to cover the indicated distance (m).

For the evaporative case with 0·5 m distancing, the median viral count accumulates rapidly to 70, then decreases as smaller droplets with lifetimes under 1 s evaporate into nuclei. Between 4·5 to 8 s, viral count increases from 33 copies up to 121 also due to the passage of a cloud of large droplets drifting close to the ground. Compared to the non-evaporative case, this increase is more significant, because large droplets tend to evaporate slowly compared to small ones (Supplementary Information), and as droplet lifetimes are increased as they shrink in size and they remain airborne for longer periods of time. In addition, the non-volatile components are conserved, so the viral loads (in copies per unit volume) in evaporating droplets are higher than the original viral loads released from the source. Upon complete droplet evaporation, viruses persist in the dry residue as airborne nuclei, which continue to pose infection risks over extended distance and time. Here, the viral counts shown in Figure 3 represents droplet transmission and excludes viruses contained in airborne nuclei.

The number of viral copies required to establish an infection in half of individuals, or Infectious Dose (ID), is wide-ranging and depends significantly on the type of virus. The ID for SARS-CoV-2 is currently unknown but expected to be small, based on how fast the virus has spread globally. Since the viral load sampled from patients on admission day is more than 17 times the median load,[39] reading off at a rescaled viral dose of ~6 (100/17) would suggest significant infection probabilities.

**Two persons separated by 1 m**

For 1 m distancing, Figure 4 shows snapshots of 2–100 μm droplet dispersion up to 10 s following a cough, side and top-down view, without (top) and with evaporation (bottom). With a Listener model as an obstacle to air flow, the aerodynamics is modified so that the droplet plume is elevated and its angle is now horizontal from the Cougher's chest level up to 10 s following the cough. The presence of the Listener at 1 m has effectively increased the dispersion range of the droplet plume, with practical consequences. Observe that the head of the Listener is engulfed by the droplet plume when it should not be at that distance (Figure 1). Also, now since air has to move around the Listener, the lateral dispersion length is now defined by the characteristic width of the Listener (shoulder to shoulder), which by inspection, results in a 30° wedge, instead of the 20° wedge previously obtained for single Cougher case (Figure 1). Further downstream, the lateral plume angle is horizontal and as before, the droplets further from the center travel faster than those near the center due to entrainment of air in the Listener's wake.

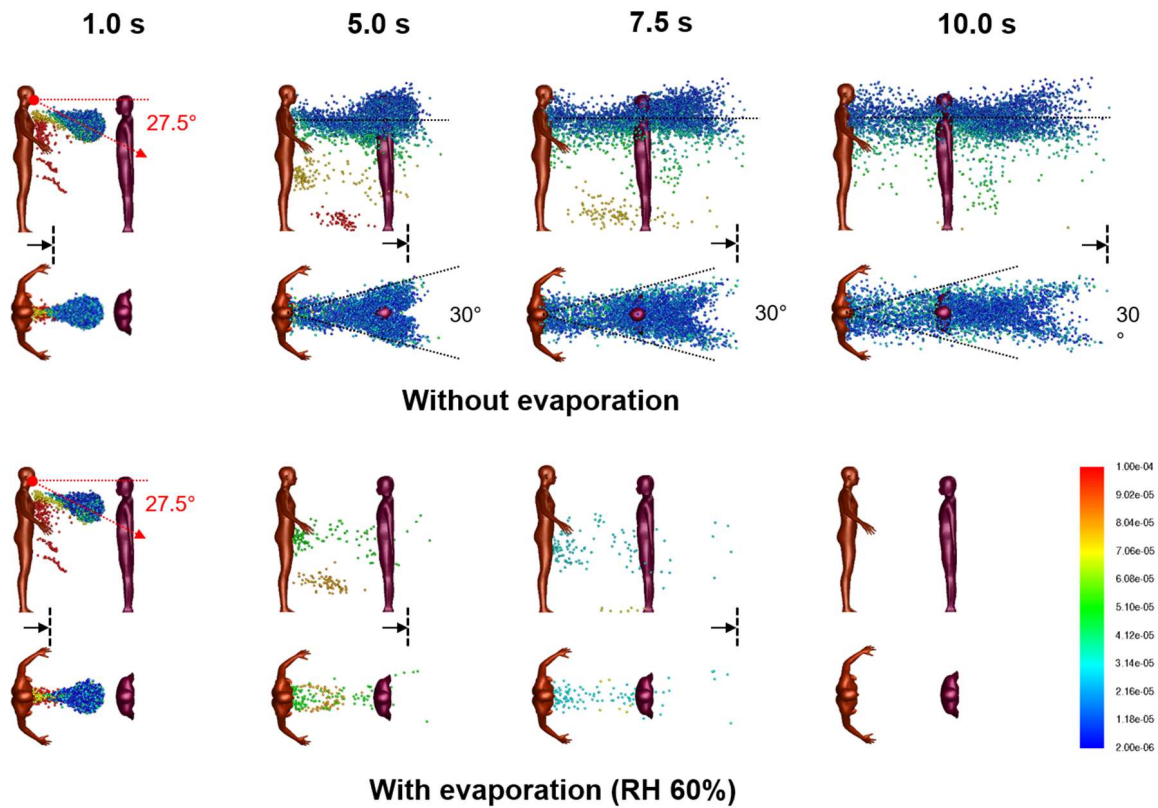

**Figure 4.** Droplet dispersion (side and top-down views) from a single cough inclined downwards at 27·5° for non-evaporative (top) and evaporative (bottom) cases at relative humidity of 60%. Listener is 1 m away facing Cougher. Color bar indicates droplet sizes (2–100 μm). Arrows represent drift markers based on a background wind speed of 0·3 m/s. Ambient air temperature is 25°C and breath temperature is 36°C. Plume is aligned horizontally from the chest up to 10 s after the cough and lateral dispersion fit a 30° forward wedge.

For the evaporative case, significant reduction in droplet counts is observed seconds from the cough (Figure 4). Unlike the non-evaporative case, the dispersion of evaporating droplets is found to be relatively unaffected by the presence of the Listener. Dispersion of droplet nuclei is expected to follow air streamlines due to their small sizes.

Figure 5 shows the air flow field with Listener 1 m away at 0·1 and 5·0 s from the cough. At the mouth level, the velocity field is initially dominated by the high velocity air jet from the cough. Both the width of the cough jet and the decay of jet speed fall within expected ranges reported in experimentally measured cough profiles.[41] At the waist level, the flow field is unaffected by the cough jet.

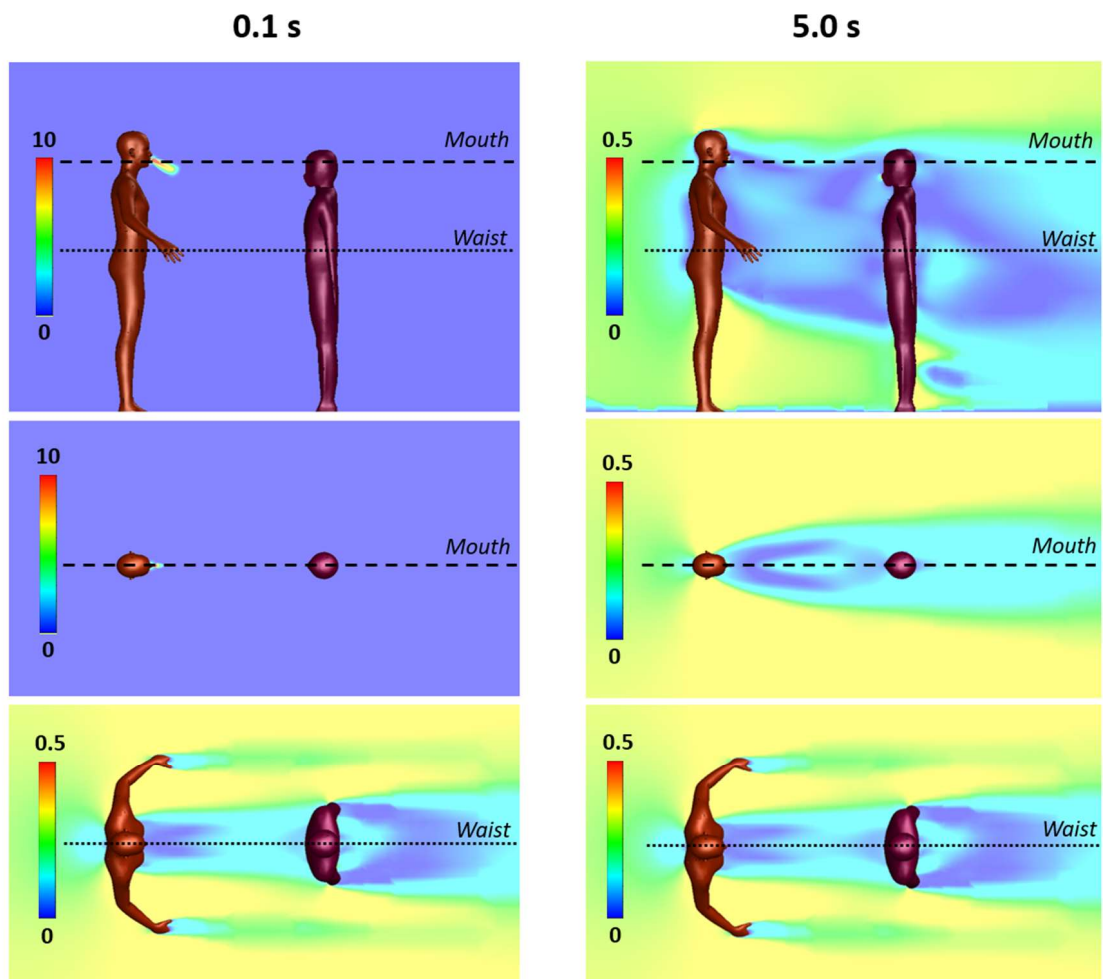

**Figure 5.** Velocity fields between a Cougher and a Listener spaced 1 m apart at 0·1 and 5 s following a single cough inclined downwards at 27·5°. Color bar indicates velocity magnitude. Top panels: side view (center plane); middle panels: top-down view (cross-section at Cougher mouth level); Bottom panels: top-down view (cross-section at Cougher waist level) reveals steady flow fields.

## Two persons separated by 2 m

For 2 m distancing, Figure 6 shows droplet dispersion up to 10 s following a cough, for both side and top-down views, without (top) and with evaporation (bottom). With the Listener now further away, the droplet plume reverts to the base case scenario (Figure 1), where it is inclined from 14° to 10° from the chest level of the Cougher up to 10 s. Closer to the Listener, however, the plume realigns horizontally as air wraps around the chest level. Top-down view shows that the lateral plume is now confined within a 20° forward wedge (instead of 30° for 1 m distancing), since the width of the Listener is smaller than the characteristic width of the wake at that distance. The chest presents a wider obstacle to droplet dispersion compared to the head (Figure 1), resulting in a droplet-free wake further downstream from the Listener. For the evaporative case, the dispersion of evaporating droplets is similarly unaffected by the Listener at 2 m.

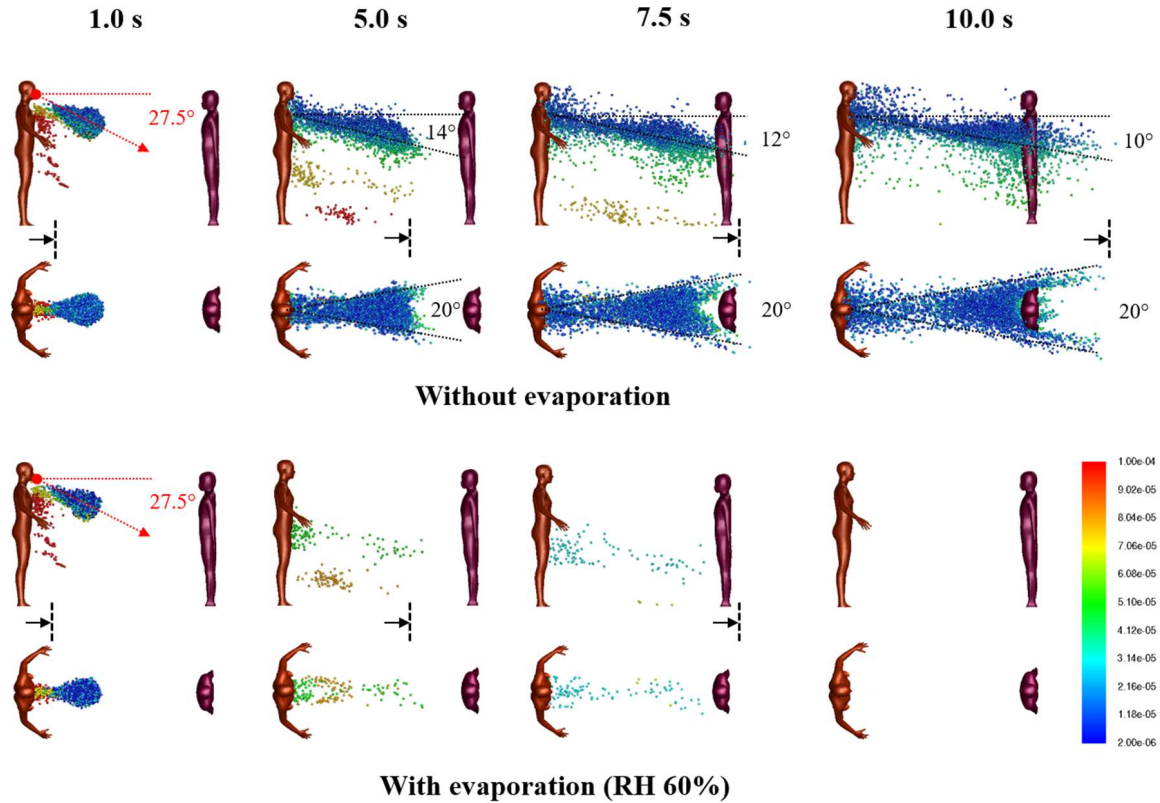

**Figure 6.** Droplet dispersion (side and top-down views) from a single cough inclined downwards at 27·5° for non-evaporative (top) and evaporative (bottom) cases at relative humidity of 60%. Listener is 2 m away facing Cougher. Color bar indicates droplet sizes (2–100 µm). Vertical lines are spaced 1 m apart and arrows are drift markers based on a background wind speed of 0·3 m/s. Ambient air temperature is 25°C and breath temperature is 36°C. Plume angles 14–10° from the chest level and lateral dispersion fits a 20° forward wedge.

**Droplet deposition and viral transmission**

Figure 7 shows that most of the droplets deposit on the body surface, and only a small fraction on the mouth region, which are inhaled or ingested. The modes for body deposition are 8–16 μm at 1 m and 16–24 μm at 2 m; the average mode for mouth and mask deposition is 4–8 μm at both distancing. Although the number of droplets deposited on the mouth (inhaled) is less than on the body, the deposition efficiency (per unit area) is substantial since the surface area of the mouth region is only 0·02% of the entire body. The droplet Stokes number ranges from $10^{-6}$ for 2 μm droplet to 0.002 for 100 μm droplet.

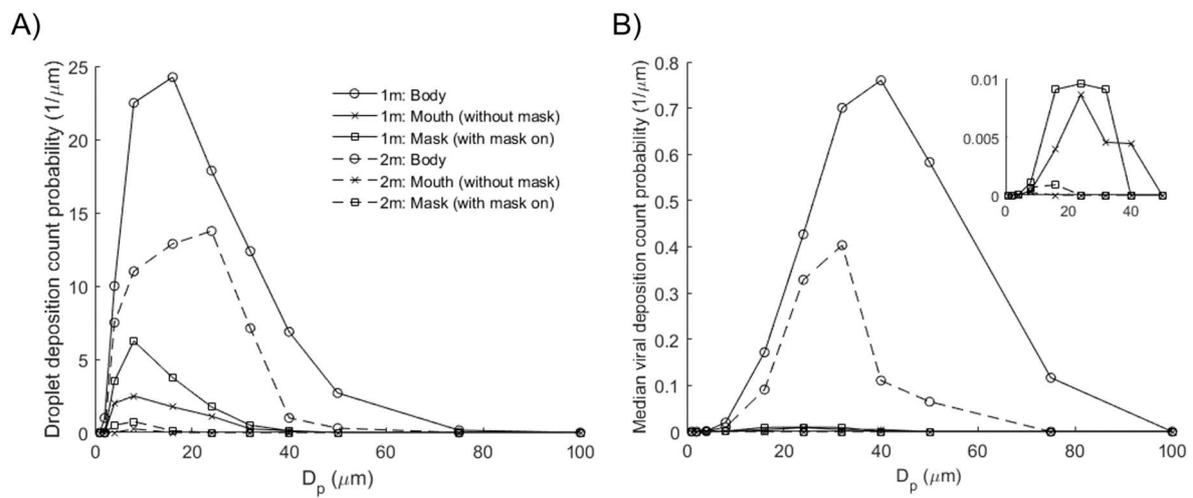

**Figure 7.** Droplet and virus deposition on the surfaces of the Listener model, including body (1·43 $m^2$), mouth (3·5 $cm^2$) and mask (288 $cm^2$). A) Droplet deposition count probability. Modes: 8–16 μm (body; 1 m), 16–24 μm (body; 2 m) and 4–8 μm (mouth/mask). B) Median viral deposition count probability (3·3×$10^6$ copies/mL)[39]. Modes: 32–40 μm (body; 1 m) and 24–32 μm (body; 2 m). Inset shows viral deposition count probabilities for mouth and mask. Mode: 16–24 μm (mouth/mask; 1 m). For 1 m distancing, the total deposited viral counts on the Listener model are 25·3 (body), 0·55 (mouth) and 0·72 (mask).

During the inhalation half cycle (Supplementary Figure S1), droplets are transported by the air intake into the mouth where they may be ingested, inhaled or exhaled. Droplet deposition on the mask is enhanced due to breathing cycles. Inhaled droplets are generally smaller (mode: 4–8 μm) in size compared to those deposited on the Listener's body (mode: 8–24).

The median viral deposition count probability based on SARS-CoV-2 viral load ($3 \cdot 3 \times 10^6$ copies/mL)[39] shows significant viral deposition on the body surface (Figure 7). Interestingly, the highest viral counts found on the body are deposited by droplets of sizes 32–40 μm at 1 m and 24–32 μm at 2 m, which is the opposite trend for droplet counts (8–16 μm at 1 m; 16–24 μm at 2 m). This is because large droplets 32–75 μm settle preferentially at distance 1–2 m, leading to significant reductions in viral counts. Total deposited viral counts are $25 \cdot 3$ viral copies on the Listener's body, $0 \cdot 72$ on mask, and $0 \cdot 55$ on mouth (inhaled). When extrapolated to Day-0 viral load,[39] the inhalation exposure increases to $9 \cdot 3$ copies, which remains lower than the Infectious Dose (ID) of most common viruses. Increasing distancing to 2 m results in significant reductions in viral deposition counts on the body and transmission risks from droplet inhalation.

## DISCUSSION

Young children may be at greater risk from droplet transmission compared to adults (Figure 1). Inspection of the droplet plume shows maximum droplet count densities at characteristic heights at $1 \cdot 2$ m for 1 m distancing, and $1 \cdot 0$ m for 2 m. A useful guideline for height-related risk would be within a height difference of 50 cm at a distance of 1 m, and 70 cm a distance of 2 m, depending on the height of the Cougher.

Most airborne droplets are 8–16 μm in diameter, but droplets with the highest transmission potential are, in fact, 32–40 μm, due to their higher viral content (Figure 2). Surgical masks are known to be effective at trapping these larger droplets, so they are recommended for use as necessary.

Our results suggest that social distancing is generally effective at reducing droplet counts across all droplet sizes. Specifically, an increase in distancing from 0.5 m to 1 m significantly reduces potential exposure to large droplets greater than 75 µm, and further increase in distancing to 2 m further halves the transmission potential across all droplet sizes. Understanding the infectious dose of SARS-CoV-2 is essential to quantitative risk assessment.

Under low humidity conditions, small droplets evaporate rapidly in a fraction of a second, whereas moderately large droplets can persist over a few seconds and remain airborne (Figure 3). Evaporated droplet residues, or nuclei, contain high viral densities and present long ranged airborne transmission risks. Smaller and lighter, these droplet nuclei contain viral residues in a compact form and could remain airborne for long period of time, projecting transmission risks over long distances. The effect of desiccation on the viability of SARS-CoV-2 found in droplet nuclei is currently not well understood.

Interesting aerodynamics come into play when a person obstructs the airflow downstream from the cough (Figure 4). Counter-intuitively, the person being 1 m away from the cough may not only fail to obstruct the plume itself, but instead enhance its dispersion range further downstream. This suggests that a 1 m distancing rule between individuals in a queue may have adverse consequences.

The airflow result for 1 m distancing also has practical consequences for face shield users. Based on our modelling results, the droplet plume for a person at 1 m follows the air flow over the chest (Figure 4), so the droplets reach the face region from a bottom-up direction which circumvents the face shield. Therefore, the face shield may be ineffective as a protection against droplets based on aerodynamic considerations. An improvement to the face shield design could be a chin plate that prevents air flow from the chest upwards; air could instead be drawn in from the back of the head around the sides.

Based on median viral load, an average of 0·55 viral copies is inhaled 1 m away from a single cough. This seemingly low figure could still be amplified through successive coughs at higher viral loads and accumulated over time. Droplets deposited on skin and clothes could still lead to secondary transmission modes such as face touching.

The present study has several limitations. First, the cough model is idealized. The droplet size distribution is based on 22 cough tests and measurements of some 3,000 droplets captured on slides[38] and the cough airflow based on 25 test subjects.[37] These sources, while representative, do not account for significant variations in coughing intensity and duration.[42] Second, droplets emitted through vocalization and sneezes are not considered in this study. Third, the height of the Cougher can affect the dispersive range of droplets, especially larger droplets which tend to settle rapidly on emission. This effectively increases the droplet transmission potential from taller infected persons. Fourth, a light prevailing wind with constant speed is assumed for indoor conditions, but wind speed significantly affects the dispersion range of droplet. Under outdoor conditions, wind speeds can reach up to several meters per second, leading to dispersion ranges that exceed currently accepted social distances even under strongly evaporative conditions.[34] Fifth, this study does not distinguish between viable and non-viable viruses, only viral count and load. Sixth, the evaporation physics used here is based on weak coupling between droplets, so droplets evaporate rapidly into smaller droplet nuclei which persists airborne. A separate study on the evaporation of droplets with non-volatile content is currently in progress.

From our data, we conjecture that the concept of "airborne" or "droplet" transmission as applied to a respiratory pathogen may be relative. Physical factors, such as wind speed and direction, interact with biological ones, such as infectious dose, to determine the likelihood with which the coughing sick infect others in their immediate vicinity.

**Supplementary Information**

**Cough characterization**

Based on established cough patterns, the standard cough is modelled as a planar jet inclined downwards at an angle 27.5° averaged between 15° and 40° angular limits of a typical cough jet.[1] The mouth opening area is 4 cm², a time-averaged value based on mouth opening data during a cough.[1] The flow velocity during cough (Supplementary Figure S1) is derived from the flow rate data based on cough tests.[1,2] At the end of the cough cycle, breathing cycle begins, modelled here as a sinusoidal function with a period of 6 s and amplitude of 0.73 m/s. Accordingly, characteristic Reynolds numbers for cough is approximately 13,000 and for normal breath 1,000. Breath temperature is assumed to be 36°C.

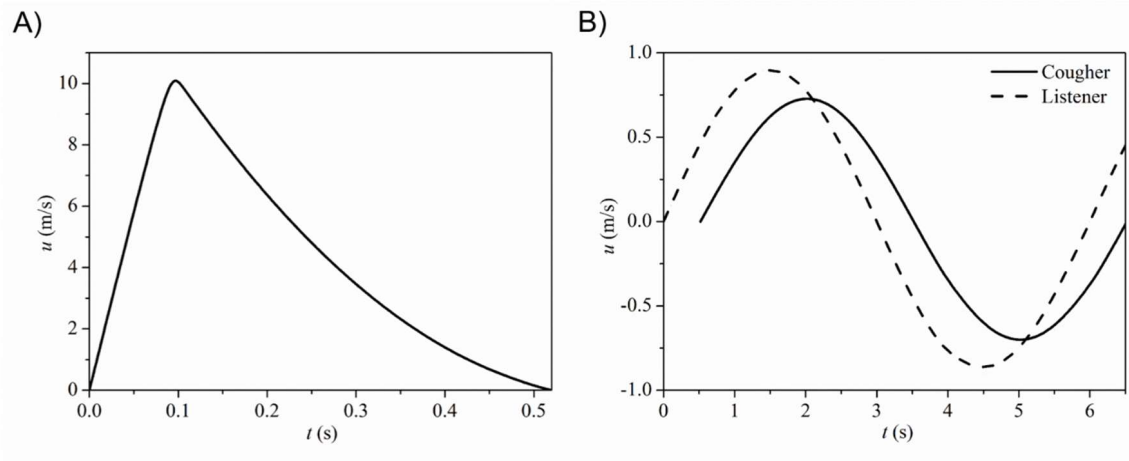

**Supplementary Figure S1.** A) Airflow velocity during modelled cough cycle. B) Breathing cycles for Cougher and Listener. For the Cougher, the breathing cycle begins at the end of the cough cycle.

The size distribution of droplets emitted in a cough is reproduced from a seminal study by Duguid[3], as shown in Supplementary Table S1. More recent measurements, carried out using different methods and under different conditions, reported size distributions that are predominantly skewed towards larger droplets (50–100 µm)[42,5]. These distributions may not sufficiently represent smaller droplet sizes emitted during a cough.[11]

| Diameter (µm) | 2 | 4 | 8 | 16 | 24 | 32 | 40 | 50 |
|---|---|---|---|---|---|---|---|---|
| Number count[*] | 50 | 290 | 970 | 1,600 | 870 | 420 | 240 | 110 |
| Median viral count (copies/droplet)[†] | $1.38 \times 10^{-5}$ | $1.11 \times 10^{-4}$ | $8.84 \times 10^{-4}$ | $7.08 \times 10^{-3}$ | $2.39 \times 10^{-2}$ | $5.66 \times 10^{-2}$ | 0.111 | 0.215 |
| Day-0 viral count (copies/droplet)[‡] | $2.38 \times 10^{-4}$ | $1.92 \times 10^{-3}$ | $1.53 \times 10^{-2}$ | 0.122 | 0.413 | 0.978 | 1.917 | 3.714 |

| Diameter (µm) | 75 | 100 | 125 | 150 | 200 | 250 | 500 | 1000 |
|---|---|---|---|---|---|---|---|---|
| Number count[*] | 140 | 85 | 48 | 38 | 35 | 29 | 34 | - |
| Median viral count (copies/droplet)[†] | 0.729 | 1.728 | 3.375 | 5.832 | 13.82 | 27.00 | 216.0 | - |
| Day-0 viral count (copies/droplet)[‡] | 12.59 | 29.84 | 58.30 | 100.7 | 238.7 | 466.4 | 3,731 | - |

[*] Reproduced from Duguid.[3]

[†] Based on median SARS-CoV-2 viral loading of $3.3 \times 10^6$ copies/mL.[42]

[‡] Based on SARS-CoV-2 viral loading from hospitalized patients on day of admission of $5.7 \times 10^7$ copies/mL.[42]

**Supplementary Table S1.** Droplet size distribution and SARS-CoV-2 viral load in a typical cough.

Based on saliva samples of hospitalized patients, the median SARS-CoV-2 viral load is found to be $3.3\times10^6$ copies/mL, with a range from $9.2\times10^2$ to $2.0\times10^8$ copies/mL.[39] Day-0 viral loading from admission day samples was reported to be $5.7\times10^7$ copies/mL and this sets the transmission risk limit for our modelled source. Assuming that salivary viral loading rates are conserved in droplets emitted during a cough, the viral content in a droplet can be determined depending on its size. Droplets with diameters greater than 100 µm neglected in this study due to their extremely fast settling rates and low dispersion potential.

For modelling purposes, droplets are generated based on discrete size distributions and injected randomly within mouth region at the start of the simulation.

## Governing equations

The problem mentioned above is governed by fluid flow, droplet movement as well droplet evaporation for the evaporation scenario. The governing equations then include Navier-Stokes equation, proper turbulent flow model as well as discrete phase movement and droplet evaporation equations, respectively.

The governing equations for fluid mass and momentum with turbulence are

$$\frac{\partial \rho}{\partial t} + \nabla \cdot (\rho \vec{u}) = \dot{m} \,,$$

(S1)

$$\frac{\partial (\rho \vec{u})}{\partial t} + \nabla \cdot (\rho \vec{u} \vec{u}) = -\nabla P + \nabla \cdot \left[ (\mu + \mu_t)(\nabla \vec{u} + \nabla \vec{u}^T) \right] - \nabla \cdot \left( \frac{2}{3} \rho \kappa \vec{I} \right) + F_m \,,$$

where $\kappa$ is the turbulent kinetic energy and $\varepsilon$ is the dissipation of turbulent energy, expressed as

$$\frac{\partial(\rho\kappa)}{\partial t} + \nabla \cdot (\rho\vec{u}\kappa) = \nabla \cdot \left(\frac{\mu_t}{\sigma_k}\nabla\kappa\right) + G_k - \rho\varepsilon,$$

$$\frac{\partial(\rho\varepsilon)}{\partial t} + \nabla \cdot (\rho\vec{u}\varepsilon) = \nabla \cdot \left(\frac{\mu_t}{\sigma_\varepsilon}\nabla\varepsilon\right) + \frac{\varepsilon}{\kappa}(C_{1\varepsilon}G_k - C_{2\varepsilon}\rho\varepsilon),$$

where $C_{1\varepsilon}$ and $C_{2\varepsilon}$ are constants 1.44 and 1.92 respectively, $\sigma_\kappa$ and $\sigma_\varepsilon$ are 1.00 and 1.3 respectively[8] and $G_k$ is the production of turbulence kinetic energy.

Eddy viscosity $\mu_\tau$ is expressed as

$$\mu_t = \rho C_\mu \frac{\kappa^2}{\varepsilon},$$

where $C_\mu$ is equal to 0.09.

The source terms in continuity and momentum (Equations S1 and S2) accounts for fluid loss via evaporation,

$$\dot{m} = \frac{\Delta m_d}{m_{d0}}\frac{\dot{m}_{d0}}{V},$$

$$F_m = \frac{1}{V} \left( \sum \left( \frac{18 \mu C_D \operatorname{Re}_D}{24 \rho_d D_d^2} (\vec{u}_d - \vec{u}) \right) \dot{m}_d \Delta t \right),$$

(S7)

where $m_{d0}$ is the mass of droplet, $\dot{m}_{d0}$ is the rate of change of droplet mass and $V$ is the control volume.

In addition to solving the flow field, species transport equations are also solved. Air is assumed to consist three main species components, i.e. $O_2$, $N_2$ and $H_2O$ vapor. The mass fraction of $O_2$ and $H_2O$ is solved by

$$\frac{\partial (\rho x_i)}{\partial t} + \nabla \cdot (\rho \vec{u} x_i) = \nabla \cdot \vec{J}_i + S_i,$$

(S8)

where $\vec{J}_i$ is the diffusive flux of species i and can be expressed as,

$$\vec{J}_i = -\left( \rho D_{i,m} + \frac{\mu_t}{Sc_t} \right) \nabla x_i - D_t \frac{\nabla T}{T},$$

(S9)

where $Sc_t$ is the turbulent Schmidt number (taken as 0.7) and $D_t$ is the turbulence diffusivity. The source term of species i is simply

$$S_i = \frac{dm_d}{dt} \frac{1}{V}.$$

(S10)

The energy conservation equation is,

$$\frac{\partial(\rho E)}{\partial t} + \nabla \cdot (\rho \vec{u} E) = \nabla \cdot \left( (\lambda \nabla T) - \sum_i h_i \vec{J}_i \right) + S_h,$$

(S11)

where E is energy,

$$E = h - \frac{P}{\rho} + \frac{\vec{u}^2}{2},$$

(S12)

h is the sensible heat,

$$h = \sum_i x_i C_{p,i} T + \frac{P}{\rho},$$

(S13)

$S_h$ is thermal source term,

$$S_h = \frac{1}{V} \left( \frac{\dot{m}_{d0}}{m_{d0}} (m_{d,in} - m_{d,out}) h_{fg} + m_{d,in} C_{pd} T_{in} - m_{d,out} C_{pd} T_{out} \right),$$

(S14)

where the subscripts in and out identifies droplets entering and exiting a control volume.

The equation of motion of a droplet (subscript d) is

$$\frac{d\vec{u}_d}{dt} = F_D (\vec{u}_d - \vec{u}) + \frac{\vec{g}(\rho_d - \rho)}{\rho_d},$$

(S15)

$\vec{u}_d$ and $\vec{u}$ are the droplet and air velocities respectively. $F_D$ is the drag force,

$$F_D = \frac{18\mu}{\rho_d D_d^2} + \frac{C_D \, \text{Re}}{24},$$

(S16)

where $D_d$ is the droplet diameter and $C_d$ is the drag coefficient (Morsi and Alexander, 1972) as a function of the droplet Reynolds number,[9]

$$C_D = c_1 + \frac{c_2}{\text{Re}} + \frac{c_3}{\text{Re}^2},$$

(S17)

$$\text{Re} = \frac{\rho D_d \left| \vec{u}_d - \vec{u} \right|}{\mu},$$

(S18)

where $c_1$, $c_1$, $c_3$ are empirical constants for spherical droplets estimates at following Reynolds number intervals,

$$c_1, c_2, c_3 = \begin{cases} 0, & 24, & 0 & 0 < \text{Re} < 0.1 \\ 3.69, & 22.73, & 0.0903 & 0.1 < \text{Re} < 1 \\ 1.222, & 1.667, & -3.889 & 1 < \text{Re} < 10 \\ 0.6167, & 46.5, & -116.67 & 10 < \text{Re} < 100 \\ 0.3644, & 98.33, & -2778 & 100 < \text{Re} < 1000 \\ 0.357, & 148.62, & -47500 & 1000 < \text{Re} < 5000 \\ 0.46, & -490.546, & 578700 & 5000 < \text{Re} < 10000 \\ 0.5191, & -1662.5, & 5416700 & \text{Re} > 10000 \end{cases}$$

(S19)

Droplet evaporation is governed by diffusive flux of droplet vapor into the air,

$$N_v = k_c \left( C_{vd} - C_{va} \right),$$

where $N_v$ is the molar evaporative flux of vapor and $k_c$ is the mass transfer coefficient. $C_{vd}$ is the saturated vapor pressure $P_{sat}$ at the droplet surface,

$$C_{vd} = \frac{P_{sat}}{RT_d} \; ,$$

(S21)

where $T_d$ is droplet surface temperature. $C_{va}$ is the partial vapor pressure,

$$C_{va} = x_v \frac{P}{RT_a} \, ,$$

(S22)

where $x_v$ is species mole fraction, $P$ and $T_a$ are the local pressure and temperature respectively. The mass transfer coefficient $k_c$ is correlated with Reynolds number and Schmidt number,[10]

$$k_c = \frac{D_{dv}}{D_d} \left( 2.0 + 0.6 \, \mathrm{Re}^{0.5} \, Sc^{1/3} \right),$$

(S23)

where $D_{dv}$ is the diffusion coefficient of vapor in the air. The mass of the droplet evolves as

$$m_d(t + \Delta t) = m_d(t) - N_v A_d M_d \Delta t \, ,$$

(S24)

where $m_d$ is the droplet mass, $M_d$ the molecular weight and $A_d$ is the surface area.

The droplet temperature is governed by thermal balance including latent and sensible heats,

$$m_d C_{pd} \frac{dT_d}{dt} = h A_d \left( T_a - T_d \right) - \frac{dm_d}{dt} h_{fg}$$

(S25)

where $h_{fg}$ is the latent heat of droplet. The convection heat transfer coefficient h is calculated with a modified Nusselt number,[11]

$$h = \frac{\lambda \ln\left(1 + B_T\right)}{D_d B_T} \left( 2 + 0.6 \mathrm{Re}_d^{0.5} \mathrm{Pr}^{\frac{1}{3}} \right),$$

(S26)

where Pr is the Prandtl number and $\lambda$ is the thermal conductivity of air. $B_T$ is the Spalding heat transfer number,

$$B_T = \frac{C_{pv}\left(T_a - T_d\right)}{h_{fg} - \frac{q_d}{\dot{m}_d}},$$

(S27)

where $\dot{m}_d$ is the droplet evaporation rate and $q_d$ is the heat energy transferred to the droplet.

**Evaporation model**

The evaporation model is verified in independent tests on 1, 10 and 100 μm droplets under relative humidity levels 0, 0.6 and 0.8 (Supplementary Figure S2). At relative humidity 0.6, a 10 μm droplet evaporates completely in 0.2 s, and a 100 μm in 2 s, which is comparable to simulation time scales, so evaporation is clearly an important factor in droplet dispersion. We compare our evaporation model against other studies and find agreement on droplet evaporation times across droplet sizes, based on Reynolds numbers, humidity and temperature gradients.[12,13]

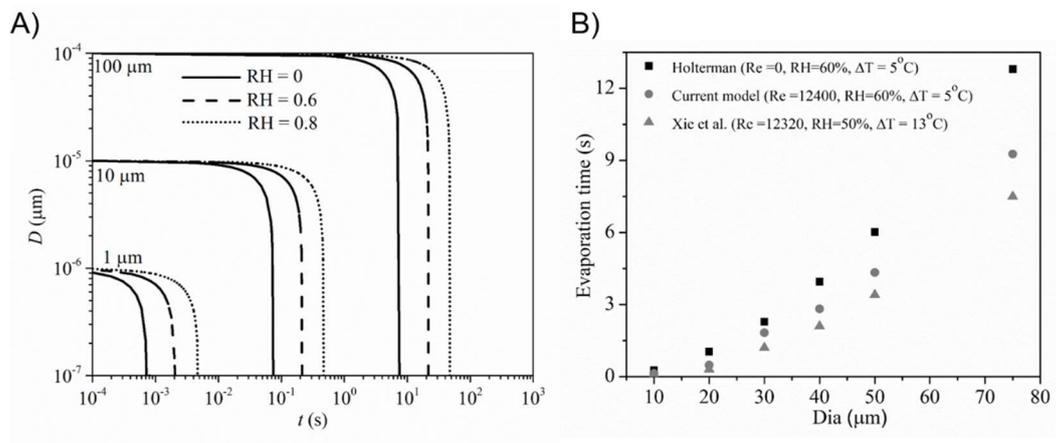

**Supplementary Figure S2.** A) Evaporation time for a droplet of diameter 1, 10 and 100 μm under relative humidity levels 0, 0.6 and 0.8. B) Evaporation times generated by present model compare well with those reported by other references.[12,13]

**Computational details**

The simulation domain is a rectangular box with dimensions of 10·1 m (length) × 7·12 m (width) × 5·4 m (height). Computational meshing is conducted directly on ANSYS FLUENT (2019R3) using polyhedral unstructured scheme. The human model is geometrically meshed as a static obstacle 1·7 m with open arms. Additional mesh refinement is implemented using double boundary layers near geometric surface of the human models. The number of mesh elements is approximately 3·9 million.

Simulations are run on standalone workstation with 88 hyper-threaded CPUs and 384G RAM. Each job is assigned 12 CPUs and 64G RAM and resolves 30 s of simulation time in approximately 24 h real time.